\documentclass[journal,10pt]{IEEEtran}
\usepackage[T1]{fontenc}
\usepackage{amsmath,bm,amsfonts,amssymb,graphicx,color,multirow,setspace,xfrac,comment,xcolor,citesort}
\usepackage{booktabs}
\usepackage{epstopdf}
\usepackage{mathtools}
\usepackage{lipsum}
\usepackage{cite}
\usepackage{multirow}
\usepackage[normalem]{ulem}
\useunder{\uline}{\ul}{}
\usepackage{float}
\usepackage{balance}
\usepackage{gensymb}
\newcommand{\tabitem}{~~\llap{\textbullet}~~}
\ifCLASSINFOpdf
\else
\fi
\begin{document}
%
\title{Six Critical Challenges for 6G Wireless Systems}
%
%
%

\author{Harsh~Tataria,~\IEEEmembership{Member,~IEEE,}~Mansoor~Shafi,~\IEEEmembership{Life Fellow,~IEEE,}\\Mischa Dohler,~\IEEEmembership{Fellow,~IEEE,}~and~Shu~Sun,~\IEEEmembership{Member,~IEEE\vspace{-25pt}}
\thanks{H.~Tataria is with Ericsson AB, Lund, Sweden (e-mail: harsh.tataria@ericsson.com).}
\thanks{M.~Shafi is with Spark New Zealand, Wellington, New Zealand (e-mail: mansoor.shafi@spark.co.nz).}
\thanks{M.~Dohler was with the Center for Telecoms Research, Department of Engineering, King's College London, London, UK. He is now with Ericsson Inc., Silicon Valley, CA, USA (e-mail: mischa.dohler@ericsson.com).}
\thanks{S.~Sun was with the Next Generation and Standards Group, Intel Corporation, Santa Clara, CA, USA. She is now with the Department of Electronic Engineering, Shanghai Jiao Tong University, Shanghai, China (e-mail: ss7152@nyu.edu).}}

%
%

\markboth{IEEE Vehicular Technology Magazine}%
{Tataria \MakeLowercase{\textit{et al.}}: Six Critical Challenges for 6G}
%



\maketitle

\begin{abstract}
A large number of papers are now appearing on sixth-generation (6G) wireless systems, covering different aspects, ranging from vision, architecture, applications, and technology breakthroughs. With cellular systems in mind, this paper presents six critical, yet fundamental challenges that must be overcome before development and deployment of 6G systems. These include: Opening the sub-terahertz (sub-THz) spectrum for increased bandwidths and the ability to utilize these bandwidths, pushing the limits of semiconductor technologies for operation within the sub-THz bands, transceiver design and architectures to realize the high peak data rates, and realizations of sub-millisecond latencies at the network-level to achieve the 6G key performance indicators. Additionally, since 6G systems will not be introduced in a green fields environment, backwards compatibility with existing systems is discussed. Where possible, we present practical solutions to realize the identified challenges.
\vspace{-7pt}
\end{abstract}


%
\IEEEpeerreviewmaketitle

\vspace{-12pt}
\section{Introduction} 
\label{Introduction}
\vspace{-2pt}
Fifth-generation new radio (5G-NR) systems are now a commercial reality. Third generation partnership project (3GPP) Releases 16 and 17 aim to serve as key enablers for the evolution of 5G-NR, capturing the inter-working capabilities of enhanced mobile broadband (eMBB), massive machine-type communication (mMTC), and ultra reliable low-latency communication (uRLLC). In parallel, a leap into the future is being taken by the international telecommunications union (ITU-T) to develop a focus group called Network 2030 \cite{ITUTNW2030}, which studies the capabilities of networks for 2030. We refer to these as the sixth-generation (6G) of wireless systems. In \cite{SAMSUNGWP1}, disruptive technologies in response to lifestyle/societal changes are predicted, which include: (1) A \emph{holographic society} where holograms/immersive reality will form a preferred means of communications; (2) \emph{Connectivity for all things} much higher than with 5G; and (3) \emph{Time-sensitive communications} where sensors form the end-points of communication.

Over the past 12 months, a large volume of literature has appeared on 6G - see \cite{ZHANG1,GIORDANI1,TATARIA1,ITUTNW2030,ITUT2,ITUT3,HONG1} and references within. Nonetheless, a rigorous discussion of the critical research challenges focused around standardization, deployment and commercial adoption of 6G systems has not been covered. This is precisely the focus of this paper. We identify six practical, yet fundamental challenges spanning the entire 6G system, which need to be overcome before development and deployment phases. Using cellular systems as our axis of exposition, we exemplify the following areas: Efficient spectrum utilization, efficient design of radio transceivers, realization of ultra low latencies (uLL), and an analysis on the constraints imposed by backwards compatibility. \footnote{We do not neglect the importance of other challenges across multiple frequency bands which may be as critical. Our derivation of challenges is primarily centered around an eventual standardization, commercialization and deployment perspective.} For each challenge, we identify practical limitations imposed by the technology and draw meaningful conclusions on eventual system-level impact.

Table~\ref{Tab:PIs} demonstrates the key performance indicators (KPIs) of 6G systems relative to 5G. For real-time holographic and immersive communication applications, data rates of $\geq$1 terabit-per-second (Tb/s) are quoted. This in turn requires the use of spectrum which can provide the needed bandwidths. As shown in \cite{ZHANG1,GIORDANI1}, a suitable candidate band lies within the range of 140-350 GHz - a.k.a. \emph{window W1} of the sub-terahertz (sub-THz) band (see Figure~4 in \cite{TATARIA1}).\footnote{{\color{black}Naturally, not all 6G services will be viable in these bands, and existing bands (microwave and millimeter-wave) will continue to play a vital role to strike the right balance between wide area coverage and optimizing peak data rates.}} Here, tens of GHz of bandwidth is available for use. However, \emph{utilizing} the available bandwidth for a working system is a challenge. We discuss the trade-offs involved in utilizing large bandwidths and provide a cautionary tale of its impact on system performance. 
\begin{table*}[!t]
\centering
\scalebox{0.68}{
\begin{tabular}{ccc}
\toprule 
\textbf{KPIs} & \textbf{5G-NR} & \textbf{6G}
\tabularnewline
\midrule
\midrule 
\textbf{Operating Bandwidth } & Up to 400 MHz for sub-6 GHz bands & Up to 400 MHz for sub-6 GHz bands\tabularnewline \textbf{(Spectrum Band Ranges)} & (band dependent) & Up to 3.25 GHz for millimeter-wave (mmWave) bands\tabularnewline
& Up to 3.25 GHz for mmWave bands & Indicative value: 10-100 GHz for THz bands \tabularnewline\hline
\textbf{Carrier Bandwidth} & 400 MHz & $\geq$ 400 MHz  \tabularnewline\hline
\textbf{Peak Data Rate} & 20 Gb/s & $\geq$ 1 Tb/s\tabularnewline 
& &(Holographic and immersive applications)\tabularnewline\hline
\textbf{User Experience Rate} & 100 Mb/s & 1 Gb/s\tabularnewline\hline
\textbf{Connection Density} & $10^6$ devices/km$^2$ (mMTC) & $10^7$ devices/km$^2$ (Connectivity for all things/ultra mMTC)
\tabularnewline\hline
\textbf{User Plane Latency} & 4 ms (eMBB) and 1 ms (uRLLC) & 10 $\mu$s to 1 ms\tabularnewline
& & (Interactive holography, immersive, and time-sensitive applications) \tabularnewline\hline
\textbf{Control Plane Latency} & 20 ms & $\leq$ 20 ms\tabularnewline\hline
 \textbf{Mobility} & 500 km/h & 1000 km/h\tabularnewline
& & Handling multiple moving platforms (terrestrial, satellites, etc.)\tabularnewline\hline
\textbf{Mobility Interruption Time} & 0 ms (uRLLC) & 0 ms (high speed mobile applications, non terrestrial networks, etc., \tabularnewline
& & and time-sensitive applications)\tabularnewline\hline
\textbf{Reliability} & 10$^{-5}$ (uRLLC)* & Up to 10$^{-7}$ \tabularnewline
& & (Holographic, immersive, and time-sensitive applications)\tabularnewline\hline
\bottomrule
\end{tabular}}
\vspace{6pt}
\caption{Anticipated requirements of 6G systems and a comparison of the 6G performance indicators relative to 5G systems. The *denotes success probability of transmitting a layer two protocol data unit of 32 bytes within 1 ms in channel quality of coverage edge. Note that the 5G KPIs are obtained from ITU-R M.2410 (Minimum Requirements for Technical Performance of International Mobile Telecommunications-2020 Radio Interfaces) and 6G KPIs are derived from \cite{ITUT2}. {\color{black}As mentioned in ITU-R M.2410, the user experience rate is one which is obtained at the 5\% point of the user throughput cumulative distribution function (CDF).}}
\vspace{-20pt}
\label{Tab:PIs}
\end{table*}

If the physical size of transmit/receive antennas is not kept constant over frequency, sub-THz bands will pose tremendous challenges to facilitate communication. This renders essential: (1) \emph{Array gain}; (2) \emph{Highly directional beamforming}. To strike the performance vs. implementation complexity ``sweetspot" of transceivers, design of analog, hybrid and digital beamforming solutions have been explored. While these are currently being optimized for 5G, a large gap - a.k.a. \emph{the terahertz gap} - is open for research. In fact, the underlaying architectures of mixed-signal circuits required in the up/down-conversion radio-frequency (RF) chains remain unsolved for such bands. Since a high degree of integration will be required, transceiver operations may be conducted with complete system-on-chip solutions. We discuss the challenges impacting efficient transceiver design and operation for sub-THz bands. 

From a semiconductor viewpoint, though complementary metal-oxide-semiconductor (CMOS) processes with a feature size of 28 nm and below can be utilized\footnote{This is since they yield an $f_{\textrm{max}}$ (maximum frequency where the semiconductor is able to provide power gain) well above 250 GHz.}, other alternatives such as silicon germanium (SiGe) bipolar CMOS (BiCMOS), high electron mobility transistor (HEMT), gallium arsenide (GaAs), and type III-V materials must be considered. This is since operation above 100 GHz requires careful conditioning of the output power, phase noise, in-phase and quadrature (I/Q) imbalance, noise figure; along with semiconductor process reliability and packaging. Therefore, a more heterogeneous base may be sought after, which can strike the right balance in optimizing the above parameters with their relative cost. We discuss challenges associated with different semiconductor technologies and outline their respective capabilities. 

Operating between 140-350 GHz will require up to an order-of-magnitude more elements relative to arrays deployed between 24.5-29.5 GHz. The design of such arrays with high radiation efficiency across wider bandwidths poses enormous challenges. While the scaling of the number of elements yields narrow beamwidths, catering for wider bandwidths is difficult. This is since the array performance at the lower edge of the band may be substantially different to higher edge. As such, steering the overall array/beamforming gain towards the user equipment (UE) serves as a major challenge. We assess when it may be likely for a system to achieve Tb/s rates and provide a discussion around its challenges. 

The sub-millisecond latencies required by the time-sensitive use cases need not only an optimized physical layer (PHY), but also higher layers. Such latencies cannot be realized by the present transport and core network architectures. As such, flattening or significant reduction of the architecture is necessary. We evaluate the contributions in the end-to-end latency and provide suggestions on how to minimize these contributions. Even if all of the above was to be done, the requirement for 6G systems to fall back to 5G and forth-generation (4G) may restrict the changes which may eventually be commercialized. Challenges related to latency and backwards compatibility are discussed across two different sub-sections.

\vspace{-9pt}
\section{Challenge \#1: Bandwidth Utilization at Sub-THz Frequencies}
\label{Challenge1:BandwidthUtilizationatSubTHzFrequencies}
The expectation of going up in carrier frequencies towards the sub-THz bands brings the implicit expectation of much higher bandwidths relative to 5G. Given the sea-level atmospheric absorption profile as a function of frequency, system operation within the first frequency window, W1, seems the most practically feasible option approaching the next decade. While this introduces much larger bandwidths, realizing radio transceivers over the entire bandwidth is almost an impossible task. This is particularly the case if one wants to maintain gain and phase uniformity at the RF front-end. Even 5G-NR systems operating in the mmWave bands have a maximum carrier bandwidth of 400 MHz. Along the same line, close proximity services are now being discussed to have 1 GHz bandwidth limitation at sub-THz bands \cite{ITURM2417-0}. This is remarkable since in the first place, the adoption to such high bands has been driven by the fact that orders-of-magnitude larger bandwidths can be utilized. Yet in practice, this rarely seems to be the case.\footnote{We make this remark with a caution, and do not take for granted the data rate gains even small increases in bandwidths bring, since its effect as a \emph{pre-log} factor in the Shannon formula is much more pronounced than the corresponding signal-to-noise ratio gain \emph{inside the log.}} 

Majority of current commercial base stations (BSs) from 24.5-29.5 GHz is made up of aggregating carriers ranging from 2 to 4, i.e., each carrier is 100 MHz wide (in case of 4 carriers). Compared to a 100 MHz carrier, the noise floor of a receiver using 1 GHz bandwidth is 10 dB higher, yielding a signal-to-noise ratio (SNR) degradation by a factor of 10. Therefore, in practice, the bandwidth of a single carrier could be limited to say 100 MHz, and higher bandwidths could be obtained by aggregating component carriers providing the needed diversity in frequency. Continuing this notion, if 5 GHz of bandwidth is desired, one needs to aggregate 50 such 100 MHz carriers. A direct consequence of this is the radio transceiver (and radiating elements) need to be in calibration across the 50 carriers. This presents a formidable challenge at sub-THz frequencies, irrespective of the antenna array architecture and size, as effects of \emph{phase noise} starts to become more pronounced. As such, maximum number of carriers and maximum operable bandwidth will be a compromise based on the ability to maintain RF front-end linearity, antenna integrated RF circuits and effective isotropic radiated power limits for safe operation. This will serve as a significant research challenge in time to come.

\vspace{-6pt}
\section{Challenge \#2: Pushing the Limits of Semiconductor Technologies}
\label{Challenge2:PushingtheLimitsofSemiconductorTechnologies}
With push towards operation in window W1, efficient analog functions handled by the RF front-end will be \emph{conditio sine qua non}. To optimize involved performance trade-offs, heterogeneity in semiconductors will be important together with three-dimensional integration and packaging having antenna ``in-package" capabilities. For digital processing, existing sub-10 nm CMOS and future technologies with logic transistors which can deliver extreme speeds at reduced supply voltages and cost are required. In analog design, a sufficiently high $f_{\textrm{max}}$ is necessary, in conjunction with consistent output signal power, quality of on-chip components, noise issues and robustness to process and temperature variations. While 28 nm or below CMOS can offer $f_{\textrm{max}}$ above 250 GHz, continuous gate-length scaling leads to deterioration of gate resistance and hence limits a further increase of $f_{\textrm{max}}$. The process options relevant to window W1 (additionally to bulk CMOS) include fully depleted silicon on insulator (FD-SOI) and fin field effect transistor (FinFET). Relative to bulk CMOS, FD-SOI yields higher performance, while FinFET primarily has advantages in gain, sub-threshold slope and in shrink capability to 5 nm. Simultaneously, process development with \emph{bipolar} semiconductor technologies of CMOS and SiGe have seen increasing adoption for beyond 100 GHz RF design utilizing 90 nm process lithography. This gives rise to extended temperature range, higher reliability, longer process lifetimes. More critically, SiGe bipolar technology yields a 4$\times$ higher breakdown voltage relative to CMOS for a given $f_{\textrm{max}}$, which is of high significance in circuits like power amplifiers (PAs) and voltage controlled oscillators to achieve low phase noise. To strike the right balance between cost and performance, SiGe-BiCMOS technologies may be the best trade-off. 

\begin{figure}[!t]
    \centering
    \includegraphics[width=7.3cm]{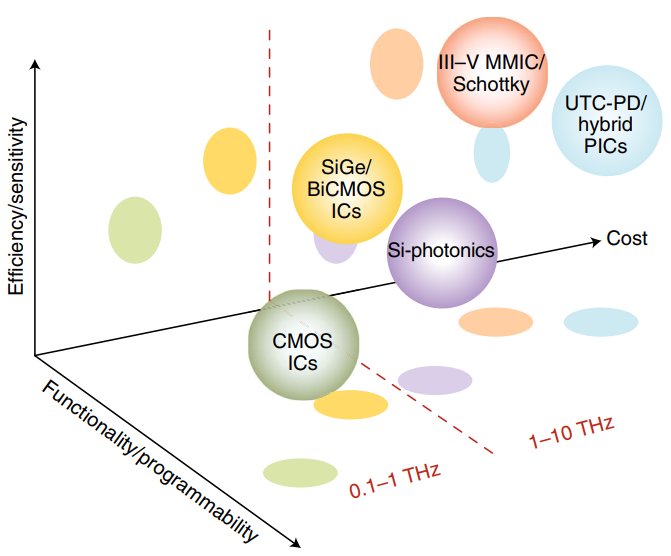}
    \vspace{-3pt}
    \caption{Qualitative illustration of technologies considering efficiency, functionality, and cost \cite{Sengupta18Nature}. IC: integrated circuit; MMIC: monolithic microwave integrated circuit; UTC-PD: uni-travelling-carrier photodiode; PIC: photonic integrated circuit. Other contenders not in the figure are described in earlier text.}
    \label{fig:Tx}
    \vspace{-18pt}
\end{figure}

To efficiently maintain high output powers within window W1, GaAs or other type III-V devices could be considered with unique properties of high sheet charge, high electron mobility and wider bandgaps. This is in contrast to the limited output powers on offer by silicon-based technologies. For efficient implementation, there are ongoing efforts to co-integrate GaAs and III-V devices with CMOS or BiCMOS \cite{TATARIA1}. This co-integration can be enabled either by using monolithic processes, where the III-V devices are placed next to CMOS in the same substrate or by employing heterogeneous integration to develop modules that  incorporate microwave elements, as well as antennas in-package. In practice, a multitude of factors need to be taken into account, such as re-configurability, efficiency, and cost, among others. Although the focus of the paper is primarily on challenges up to window W1, we present a wide range of semiconductor options which could be utilized over a much wider band than just W1 for detailed understanding. Figure~\ref{fig:Tx} demonstrates a \emph{qualitative} illustration of the relative trade-offs between going up in frequency vs. complexity and cost of contending technologies. Readers may refer to Table I in \cite{Sengupta18Nature} for a summary of \emph{quantitative} analysis on some relevant performance metrics.

\vspace{-7pt}
\section{Challenge \#3: Integrated Transceiver Design at Sub-THz Frequencies}
\label{Challenge3:IntegratedTransceiverDesignatSubTHzFrequencies}
Irrespective of beamforming type, a transceiver architecture for cellular systems typically comprises of a receiver with a low-noise amplifier (LNA), I/Q mixer, and local oscillator (LO) sequence with frequency multipliers. The transmitter is equipped with a similar architecture, where the mixer is followed by a PA. The achievable \emph{receiver noise figure} and \emph{transmitter efficiency} could be severely degraded at sub-THz relative to lower frequencies. 

For instance, operation at 250 GHz with 10000 elements will need array sizes of just 6 cm $\times$ 6 cm with elements spaced half-wavelength apart (approximately 0.6 mm). As such, interface electronics must have the same size to minimize length of interconnects, constituting an important challenge. Each ``chip" must then feature multiple transceivers; e.g., a 6 mm $\times$ 6 mm could have 10 transceivers, in which case 1000 chips would be needed for 10000 elements. The elements should be integrated on-chip removing chip-to-carrier interface losses at the expense of lower antenna efficiency. {\color{black}Here, two important challenges arise: \emph{(1) power generation,} and \emph{(2) heat dissipation.} Naturally, the most critical component at the transmitter for power gain is the PA, which will typically have <15\% efficiency at sub-THz bands using multi-carrier waveforms. Output powers between 10-15 dBm can be anticipated for PA-integrated chipsets. As such, PAs will need to operate with high back-off powers and will need to be driven close to their 1 dB compression points to maximize output power. This comes at the expense of non-linearities caused by inter-modulation products, important for out-of-band characterization and spectrum co-existence. Due to this, power dissipation from the PA/transceiver becomes a challenge. Since the overall form factor is greatly reduced, the total area for heat dissipation is significantly less. If each transceiver consumes a modest 50 mW, the total power consumption of the array will be 0.5 kW, having defining consequences on the system not being continuously active. This is critical for energy efficiency considerations.} Furthermore, generation of coherent/low-noise LO signals serves as a critical challenge. Here, a decentralized solution based on \emph{local} phase locked loops is essential to avoid generating a centralized signal and to de-correlate phase noise. 

Complete realization of sub-THz systems will suffer from the lack of low-loss/cost RF interconnects. Here, conventional chip packaging techniques are inappropriate. The issue of packaging mixed-signal circuits lies in the transition across the RF circuit ports into/out of a waveguide, often via bond-wires. Nonetheless, bond-wires within window W1 contain large reactance making them difficult for impedance matching. Scaling down the size of bond-wires presents severe issues in repeatability and increased manufacturing costs. Instead, a waveguide transition \emph{within the circuit} may be a better alternative, at the cost of bandwidth reduction. This is since modes propagating between the waveguide and the cavity need to be suppressed for retaining circuit performance. This will yield more compact integration of antennas with PA/LNA, mixers, and modulators \cite{GUNARSSON1}. In any case, finding the right balance in terms of transceiver efficiency/integration, cost, packaging and manufacturing defects is critical for successful adoption of transceivers to sub-THz bands. Furthermore, on the signal processing side, challenges such as those involved in beam management for sub-THz systems also deserve further attention, especially in light of the initial results in \cite{TAN1}.

\vspace{-8pt}
\section{Challenge \#4: Achieving Tb/s Rates in Reality}
\label{Challenge4:AchievingTb/sRatesinReality}
Operation between 140-350 GHz will require unprecedentedly massive antenna arrays. Designing such arrays that operate with high radiation efficiency poses fundamental challenges in the design of underlaying RF feed networks to support the GHz-wide bandwidths. In the context of phased arrays, while the excessive scaling of the number of elements will result in narrower beamwidths; as the bandwidth is scaled up, array performance at the lower and higher edges of the band will be substantially different.  This results in \emph{distortion} of the mainlobe radiation pattern, as well as increased powers in the sidelobes. In addition, the issue of beam squinting will be a challenge to overcome, particularly as the array is driven with wideband waveforms having high peak-to-average-power ratios. With such refined beam resolution, \emph{steering} the overall array/beamforming gain towards a UE becomes excessively difficult. Since elements will be tightly integrated on-chip, maintaining the array's gain and phase response (and suppressing the effects of antenna/circuit coupling) comprises a formidable challenge. The sub-THz propagation channel is anticipated to have \emph{low rank}, in which case distributed arrays may serve as a way forward to provide rank increase and lead to an increase in the achievable rates. {\color{black}Proposals to improve the composite channel rank and coverage with reconfigurable intelligent surfaces (RISs) have also been made in research papers. Nevertheless, their commercial deployment will be hindered by several practical challenges \cite{TATARIA1}: Successful operation of RISs will require transport links of unspecified bandwidth between cellular BSs and each RIS. The required transport bandwidth largely depends on the number of elements at the RISs, the number of control bits per-element, and their refresh rates. The added transport links will increase the control plane latency of the overall system. For multiuser operation, the transport network capacity would need to scale further with the number of UEs served, and the required number of surfaces “per-cell” remains unclear.}
\begin{figure}[!t]
   \vspace{-10pt}
    \centering
    \includegraphics[width=9.5cm]{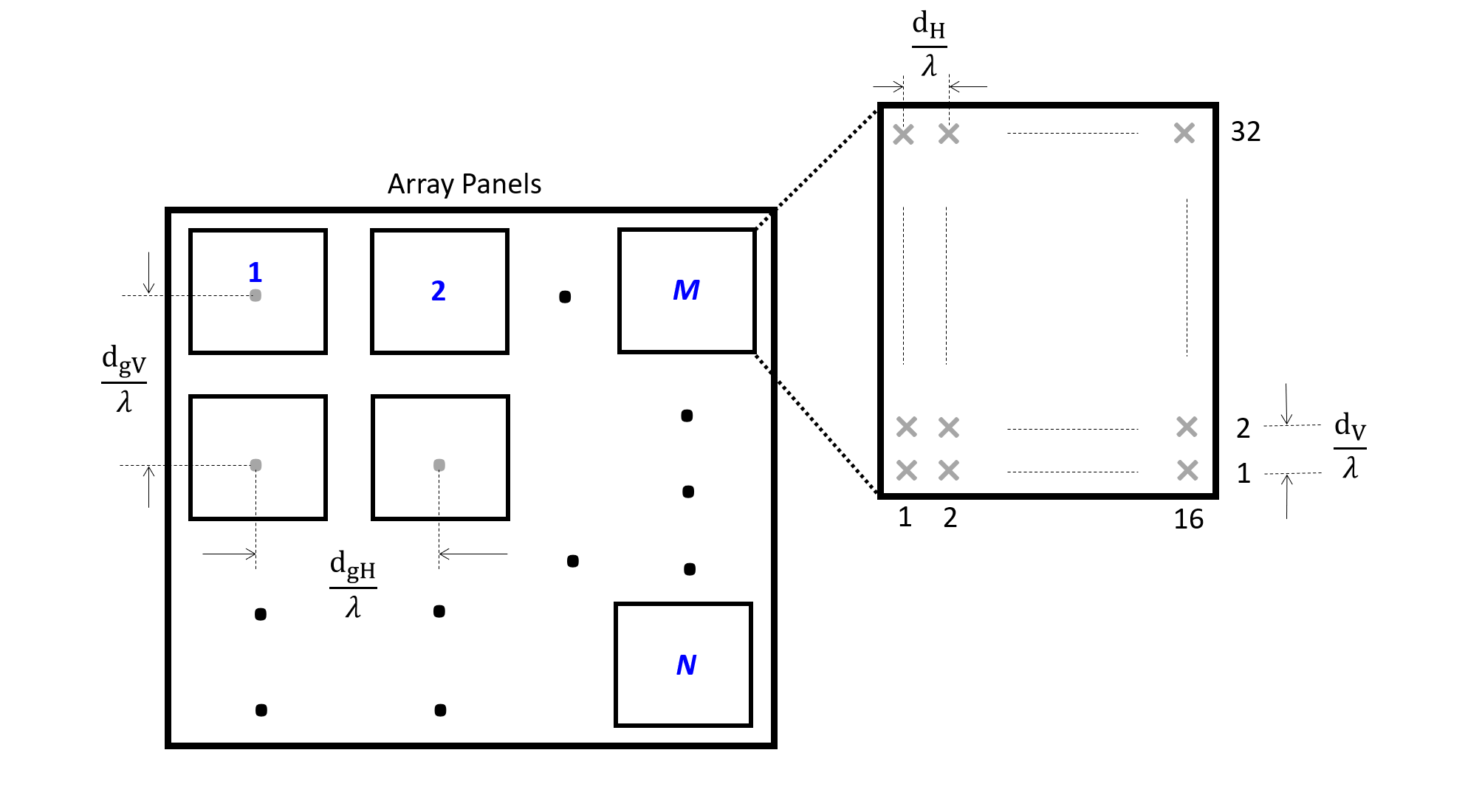}
    \vspace{-20pt}
    \caption{An example of an ultra massive multiple-input multiple-output (MIMO) array, consisting of $M\times{}N$ sub-panels, where each sub-panel has 32$\times{}$16  cross-polarized elements. The cross-polarized antenna elements in a sub-panel are spaced by $d_H/\lambda$ and $d_V/\lambda$, where $\lambda$ is the operating wavelength. On the other hand, the panels are spaced by $d_{gH}/\lambda$ and $d_{gV}/\lambda$, respectively \cite{TATARIA1}.}
    \vspace{-9pt}
    \label{fig:UltraMassiveMIMOArraysofSubArraysConcept}
\end{figure}
\begin{figure}[!t]
    \centering
    \vspace{-7pt}
    \includegraphics[width=8.8cm]{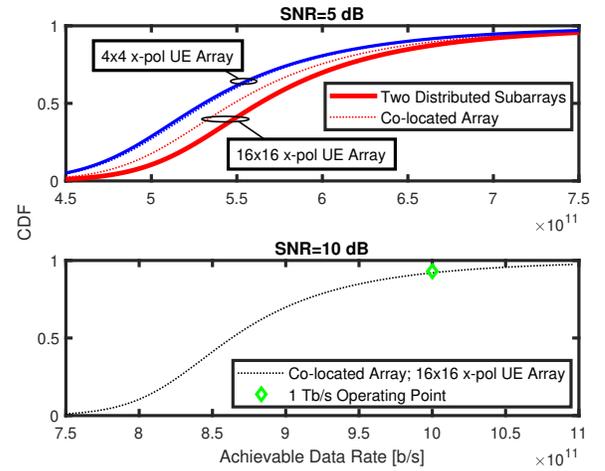}
    \vspace{-20pt}
    \caption{Single-user ultra massive MIMO data rate CDFs with 4096 BS antennas spread across four arrays (2$\times{}$2 configuration) serving a UE with 16 antennas over 10 GHz bandwidth. For both BS and UE arrays, azimuth and elevation spacing was set to 0.5$\lambda$ and 0.7$\lambda$, with per-element patterns from \cite{3GPPTR38901}. The impulse responses were generated from \cite{PRIEBE1}. The top sub-figure compares the relative data rate performance with a 4$\times{}$4 and 16$\times{}$16 cross-polarized UE arrays for co-located and two distributed sub-arrays at SNR=5 dB. The bottom sub-figure depicts the data rate CDF for a 16$\times{}$16 cross-polarized UE array at SNR=10 dB. The 1 Tb/s operating point is shown.}
    \label{fig:CapacityCDFs}
    \vspace{-17pt}
\end{figure}

To assess whether it may be possible to achieve Tb/s rates, we carry out a single-UE example. We consider two cases: In the first, we configure a co-located array of sub-arrays containing 4096 elements within a 2$\times{}$2 planar configuration. Each sub-array consists of the structure depicted in Figure~\ref{fig:UltraMassiveMIMOArraysofSubArraysConcept} with 32$\times{}$16 cross-polarized elements. Note that Figure~\ref{fig:UltraMassiveMIMOArraysofSubArraysConcept} shows a $M\times{}N$ array of sub-array for understanding purpose. The co-located array is placed at the origin of a circular cell with radius 50 m. In the second case, two arrays were distributed across opposite sides of the cell, retaining equal number of total elements. The UE employs either a 4$\times{}$4 or 16$\times{}$16 cross-polarized array in a planar fashion.  Propagation channels were generated over 140-150 GHz following \cite{PRIEBE1}, as there are no completely established channel models above 150 GHz. Figure~\ref{fig:CapacityCDFs}, top sub-figure, predicts the data rate CDFs for both cases and UE arrays at SNR=5 dB. Several trends can be observed: (1) Due to rank increase, distributed sub-arrays yield higher data rates relative to the co-located array. (2) Increasing the UE array size gives rise to a larger rate difference between distributed and co-located deployments. (3) Irrespective of the case, for CDF > 0.85, smaller difference in performance was observed due to similarity of high SNR channels. The bottom sub-figure shows co-located performance with 16$\times{}$16 UE array at SNR=10 dB, where the 1 Tb/s operating point is depicted at CDF=0.95. With this in mind, one can readily question how such high data rates can be achievable in practice while considering the aforementioned transceiver constraints. Furthermore, how will the gains scale for multiuser systems? These are important questions which need to be answered -  potentially using artificial intelligence (AI)-based solutions rather than the deterministic model-based approaches. Reference \cite{HOYDIS1} points to the challenges in integrating AI techniques to a subset of transceiver operations, such as channel estimation, symbol de-mapping and/or receiver equalization. Nevertheless, even if the AI-driven components are small, several important problems such as model updates and data explainability may need to be overcome, with fully integratable hardware accelerators into the transceiver processing flow to manage the computational overheads.

\vspace{-9pt}
\section{Challenge \#5: Realization of Global Sub-Millisecond Latencies} 
\vspace{3pt}
\label{Challenge5:Realization of Sub-Millisecond Latencies}
The importance of low latencies has been highlighted with the introduction of uRLLC in 5G. The latency KPIs for 6G are given in Table~\ref{Tab:PIs}. 
Making the user plane latency  10 $\mu$s or even 1 ms is not a trivial task. 
In this section, our end-goal is to establish if \emph{6G can enable something revolutionary}, i.e.,~a truly low latency internet which is operational across the world and not only over a localized area.

The latency contributions at PHY and RAN, as well as possible solutions to minimize these locally, have been well explored in 5G and now 6G. However, no considerations have been given to minimizing latency globally which is paramount to a uLL internet! Discussed in more detail below, the global end-to-end latency is determined by:
\begin{itemize}
\item Transport network; 
\item Edge-cloud capabilities; and
\item User applications. 
\end{itemize}

Indeed, 6G gives us the opportunity to systematically re-think the cellular architecture, and remove legacy designs, such as the \emph{transport network}. 
Keeping the traditional transport network does not allow uLL applications, such as augmented reality. Given the internet  has evolved in terms of transport capacity, one can imagine 6G to lease available transport fibre, virtualize all core network functions and focus squarely on the design of RAN and air interface. For instance, creating trusted breakout points close to the RAN and saving 2$\times$500 km transport fibre, yields a latency saving of more than 3 ms.

Another constituent to end-to-end latency is the design of the \emph{edge cloud}. Indeed, the edge cloud will be vital in enabling a true uLL internet by overcoming the fundamental limits imposed by the speed of light. As an example, the fiber connection between London and Los Angeles imposes a delay of more than 40 ms. Applications which require latencies well below that will require special provisioning through predictive AI~\cite{HOYDIS1}. 6G will thus need to transit from model-mediated telepresence systems to AI-mediated telepresence systems. The former are known to stabilize links with 120 ms latency difference; whilst the latter will achieve the same stability and will be able to adapt to the specific use case~\cite{HOYDIS1}. 

Last but not least, \emph{applications} need to be geared towards uLL operation. For instance control loops, cyber security primitives, video/audio compression algorithms need to be adapted to low delays. Here, fundamentally novel approaches to application design will be critical. For instance, the current video codec latency of more than 60 ms needs to be reduced to virtually zero if the 6G end-to-end latencies across the globe are to be achieved. 5G is able to achieve millisecond latency within a manufacturing facility but not across large geographies. With above proposed design changes, 6G will be able to offer millisecond end-to-end latency globally. Indeed, a multimedia link between continents with AI-mediated telepresence systems and an optimized application layer reduces a typical latency of >100 ms to virtually zero, thus enabling unprecedented applications.

\vspace{-9pt}
\section{Challenge \#6: Breaking the Barriers of Backwards Compatibility}
\vspace{-1pt}
\label{Challenge6:BreakingtheBarriersofBackwardsCompatibility}
When 6G systems will be deployed, they will co-exist with 5G and 4G systems. For seamless UE experience, backwards compatibility with previous generation systems will be essential. Below, we provide examples of backwards compatibility requirements, and analyze how 6G systems can be deployed. 

\begin{figure}
    \centering
    \vspace{-2pt}
    \includegraphics[width=7.8cm]{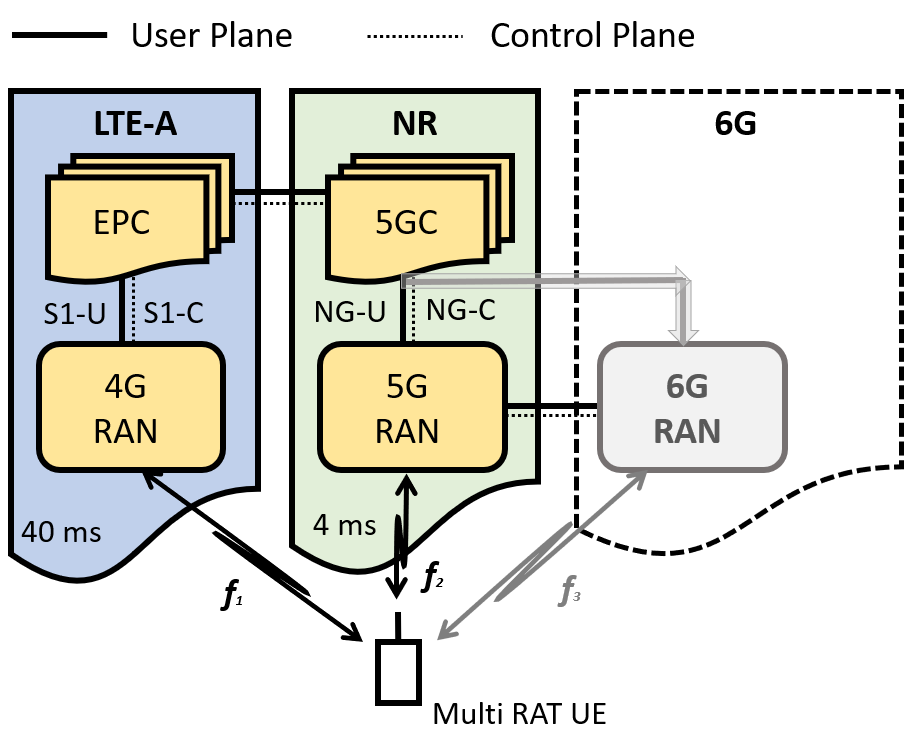}
    \vspace{-13pt}
    \caption{Possible 6G system architecture in a non-stand alone configuration co-existing with 4G and 5G-NR systems. One can note that the 6G cell will be anchored on a 5G-NR mastercell, yet the user plane traffic follows a local split approach as shown. The multi RAT UE has the capability to receive multiple frequency bands, denoted as $f_1$, $f_2$, and $f_3$. The term ``EPC" denotes ``evolved packet core", while the terms ``S1-U/C" and ``NG-U/C" define the 3GPP standardized S1 and NG interfaces for User/Control plane traffic/signalling. The interfaces between the EPC and 5G and NSA, SA operations are described in 3GPP TS 23.501 ``System Architecture for the 5G system (5GS)".}
    \label{fig:ArchitectureOptions}
    \vspace{-5pt}
\end{figure}

\begin{figure}
    \centering
    \includegraphics[width=7.8cm]{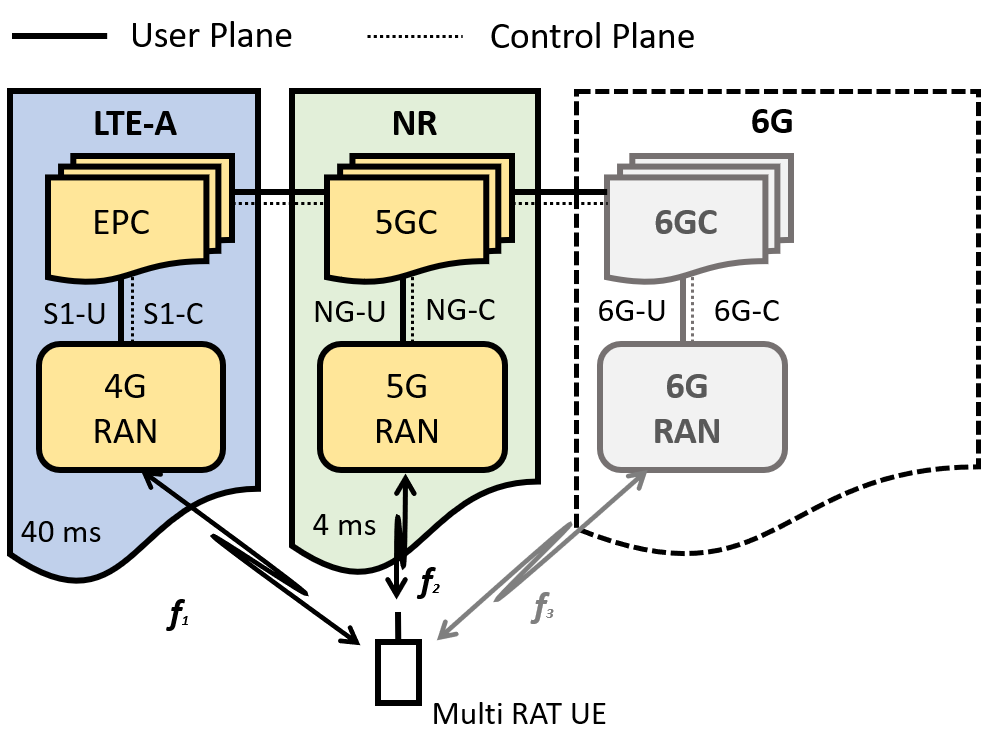}
    \vspace{-10pt}
    \caption{Possible 6G system architecture in a SA configuration co-existing with 4G and 5G systems. The control plane and user plane traffic is carried on the proposed 6G-U and 6G-C interfaces which will be capable of realizing sub-millisecond latency.}
    \label{fig:ArchitectureOptions2}
    \vspace{-10pt}
\end{figure}

In order to evaluate backward compatibility, we need to consider how the 5G and 6G core networks (5GC and 6GC) are interconnected; yet there is no \emph{current}  standardization of the 6G core in 3GPP. Assuming multi radio access technology (RAT) UEs are available, and the principles of ENDC (enhanced UTRA dual-connectivity) are adopted (as currently used for 5G-NR), 6G could  be introduced initially as a non-stand alone system, similar to current 5G deployments, then a 6G BS would act as a secondary cell to a master 5G-NR cell, and a 6G call is anchored on 5G cell, yet the user plane traffic follows a split bearer approach at the packet data convergence protocol (PDCP) layer. This architecture is depicted in Figure~\ref{fig:ArchitectureOptions}. A 6G call outside the 6G coverage area will remain anchored on a 5G master cell and will be served by the 5G cell. For this to happen, the interfaces between the 5G and 6G RANs and between the 6G RAN and the 5G core will need to be defined. This serves as a big challenge in standardization. Alternatively, 6G could be introduced as a SA, in which case a 6G call is anchored on the 6G BS and call continuity in case of no 6G coverage is via connectivity between the 5G and 6G core networks. This is illustrated in Figure~\ref{fig:ArchitectureOptions2}. Here, again the user plane and control plane interfaces between the 5G and 6G cores are yet to be defined. Irrespective of approach, a very close inter-working and backwards compatibility with previous generation systems will be the foundation of 6G. 

\begin{figure}[!t]
    \centering
    \includegraphics[width=9.3cm]{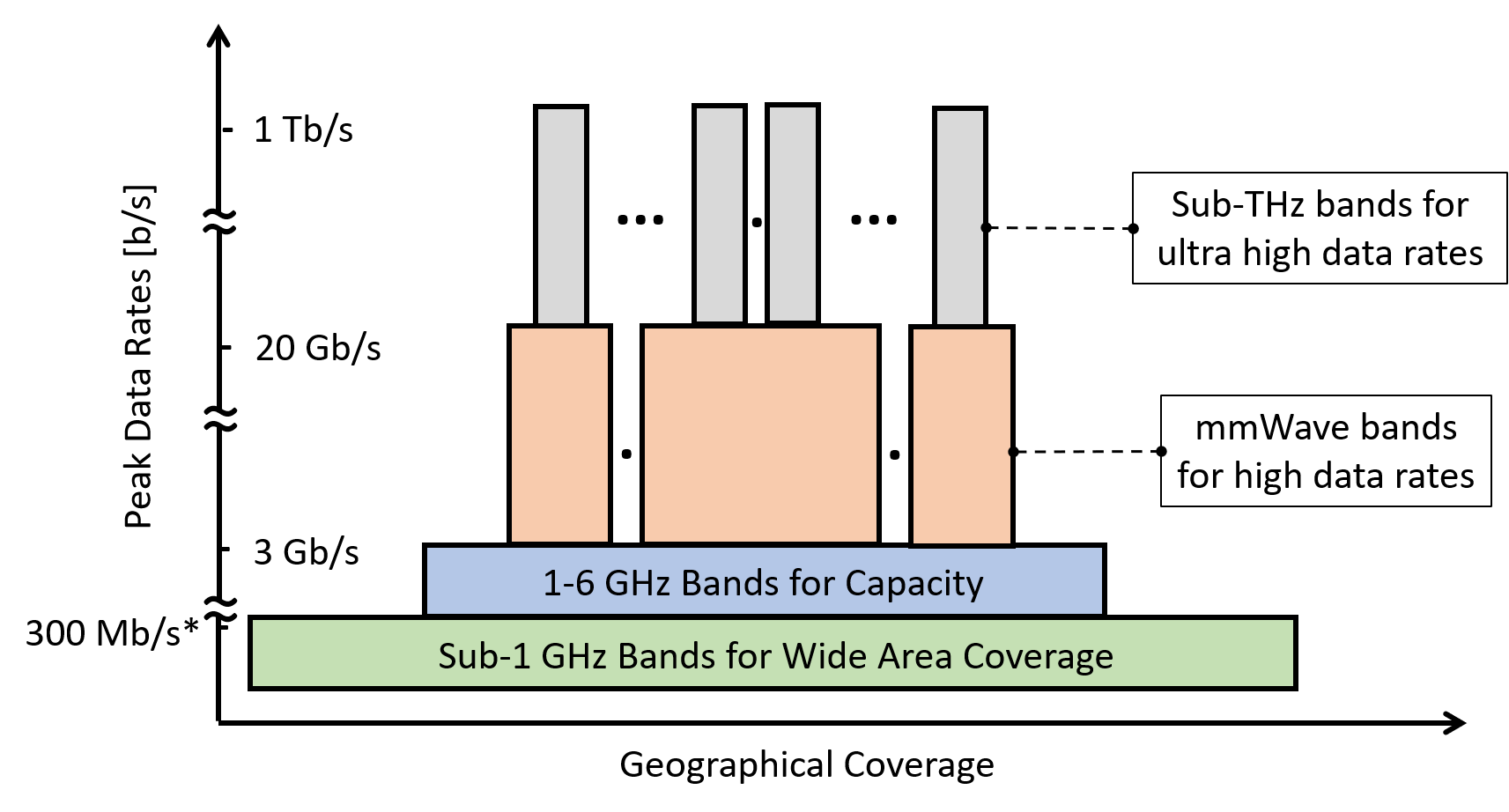}
    \vspace{-20pt}
    \caption{{\color{black} 6G operation across a wide range of frequency bands/ranges starting from sub-1 GHz to sub-THz. For each band, geographical coverage and the corresponding peak data rates are depicted. *A typical peak rate of 300 Mb/s is quoted for 20 MHz bandwidth and 4 MIMO layers, while 3 Gb/s (100 MHz bandwidth) and 20 Gb/s are quoted from ITU-R M.2410, both with the same spectral efficiency.}}
    \label{fig:6GFrequencyBands}
    \vspace{-15pt}
\end{figure}

As for the PHY, the impact of backwards compatibility on the choice of OFDM with scalebale numerologies has already been discussed earlier. This will be a key aspect if spectrum can be partly re-farmed to 6G, to enable co-existence. {\color{black}Figure~\ref{fig:6GFrequencyBands} depicts the role of a wide range of frequency bands for 6G, from sub-1 GHz to sub-THz. While the lower bands are naturally well suited to provide coverage ubiquity, the higher bands will exist in highly localized areas of varying geographical extent for providing ultra high data rates for some 6G services.} Table~\ref{Tab:WaveformComparisons} demonstrates the pros and cons of a wide range of single carrier and multi carrier waveforms and relates the different waveform types to their suitability of backwards compatibility from 6G to 5G and 4G, respectively. All acronyms used in the table are defined as in \cite{TATARIA1}.

\begin{table*}[!t]
\centering
\scalebox{0.62}{
\begin{tabular}{cccc}
\toprule 
& \hspace{220pt}\textbf{Multi Carrier Waveforms} & & 
\tabularnewline
\midrule
\midrule 
\textbf{Waveform Type} & \textbf{Pros} & \hspace{-80pt}\textbf{Cons} & \textbf{Backwards Compatible?}
\tabularnewline
\midrule
\midrule
\textbf{CP-OFDM} & \hspace{59pt}\tabitem Simpler frequency domain equalization & \hspace{60pt}\tabitem High PAPR and out-of-band-emissions (OOBE) & Yes\tabularnewline
& \hspace{22pt}\tabitem{Flexible frequency assignment} & \hspace{-13pt}\tabitem Hard coded cyclic prefix (CP) & \tabularnewline
& \hspace{35pt} \tabitem Lower implementation complexity & \hspace{5pt}\tabitem Poor performance in high mobility & 
\tabularnewline
&\hspace{8pt}\tabitem Simpler MIMO integration & \hspace{-16pt}\tabitem Stricter synchronization limits & \tabularnewline\midrule
\textbf{W-OFDM}& \hspace{-44pt}\tabitem Lower OOBE & \hspace{-42pt}\tabitem Poor spectral efficiency & No\tabularnewline
& \hspace{40pt}\tabitem Lower implementation complexity &  \hspace{-60pt}\tabitem Poor bit error rates & \tabularnewline
& & \hspace{-2pt}(Depending on window type) & \tabularnewline\midrule
\textbf{OQAM-FBMC}& \hspace{25pt}\tabitem Optimal frequency localization & \hspace{-35pt}\tabitem Challenging pilot design & No \tabularnewline
& \hspace{-3pt}\tabitem High spectral efficiency  & \hspace{58pt}\tabitem No resilience to inter-symbol  interference (ISI) & \tabularnewline
&\hspace{35pt}(Due to no guard band or CP) & \hspace{-2pt}\tabitem High implementation complexity & \tabularnewline
& \hspace{60pt}\tabitem Suitable for asynchronous transmission & \hspace{-32pt}\tabitem High power consumption  & \tabularnewline
& \hspace{48pt}\tabitem Suitable for high mobility use cases & & \tabularnewline \midrule
\textbf{F-OFDM}& \hspace{15pt}\tabitem Flexible filtering granularity & \hspace{-2pt}\tabitem High implementation complexity & No \tabularnewline 
&\hspace{18pt}\tabitem Better frequency localization & & \tabularnewline
& \hspace{-20pt}\tabitem Shorter filter length & & \tabularnewline
& \tabitem Compatible with MIMO & & \tabularnewline\midrule
\textbf{GFDM}& \hspace{11pt}\tabitem Reduced PAPR on average & \hspace{21pt}\tabitem Higher latency due to block processing & No  \tabularnewline
& \hspace{28pt}\tabitem Superior frequency localization & \hspace{-12pt}\tabitem Challenging MIMO integration & \tabularnewline
& \hspace{-37pt}\tabitem Flexible design & \hspace{-4pt}\tabitem High implementation complexity & \tabularnewline\midrule
\textbf{UFMC}& \hspace{-5pt}\tabitem Well localized filtering & \hspace{-60pt}\tabitem No immunity to ISI & No \tabularnewline
& \hspace{68pt}\tabitem Shorter length relative to sub-carrier size & \hspace{-38pt}\tabitem High receiver complexity & \tabularnewline
& MIMO compatibility & & \tabularnewline \midrule
\textbf{OTFS}& \hspace{65pt}\tabitem Ability to handle high Doppler channels & \hspace{-5pt}\tabitem Higher implementation complexity & No \tabularnewline 
& \hspace{65pt} \tabitem Exploit frequency dispersion for diversity & \hspace{10pt}\tabitem Sub-optimal equalization architectures & \tabularnewline
& \hspace{1pt} \tabitem Efficient UE multiplexing & & \tabularnewline
\midrule
& \hspace{220pt}\textbf{Single Carrier Waveforms} & & \tabularnewline
\midrule\midrule
\textbf{CP-DFT-s-OFDM} & \hspace{20pt}\tabitem All advantages of CP-OFDM & \hspace{-100pt}\tabitem High OOBE & Yes (for uplink)\tabularnewline 
&\hspace{-45pt}\tabitem Lower PAPR & \hspace{-88pt}\tabitem Hard-coded CP & \tabularnewline
& & \hspace{-5pt}\tabitem Strict synchronization requirements& \tabularnewline\midrule
\textbf{ZT-DFT-s-OFDM} & \hspace{-10pt}\tabitem Flexible guard interval & \hspace{-58pt}\tabitem Extra control signaling
 & No \tabularnewline
 & \tabitem Higher spectral efficiency & \hspace{-48pt}\tabitem Limited link performance & \tabularnewline 
 & \hspace{83pt}\tabitem Lower OOBE compared to CP-DFT-s-OFDM
 & \hspace{-15pt}(for higher-order modulation) & \tabularnewline\midrule
\textbf{UW-DFT-s-OFDM} & \tabitem Optimal spectral efficiency  & \hspace{-30pt}\tabitem All Cons of ZT-DFT-s-OFDM & No \tabularnewline 
 & \hspace{-5pt}\tabitem Lowest OOBE and PAPR & \hspace{-20pt}\tabitem High implementation complexity & \tabularnewline 
\hline\bottomrule
\end{tabular}}
\vspace{10pt}
\caption{Pros and cons of contending air interface waveforms for 6G wireless systems.}
\vspace{-20pt}
\label{Tab:WaveformComparisons}
\end{table*}


\vspace{-12pt}
\section{Conclusions} 
\label{Conclusions}
\vspace{-2pt}
Naturally, there are significant challenges that must be overcome before we can think of 6G deployments. The realization of Tb/s data rates need large bandwidths. If the 5G experience is a guide, maximum carrier bandwidths are limited to 400 MHz. This leads to aggregating large numbers of carriers to create higher total bandwidths. Building radios and associated RF circuits at sub-THz bands presents significant challenges. Therefore, finding the right balance between transceiver efficiency/integration, packaging, and cost is a key issue. Assuming transceivers can be built, achieving Tb/s rates requires ultra massive arrays. Given the low channel rank, distributed arrays will be the way forward. Time-sensitive communications need sub-millisecond latencies. We articulate an approach to achieve low latencies by addressing contributions from the air interface, RAN, and transport networks. 6G will not be introduced in green field  environments; we present a 6G system architecture and demonstrate how they will be backwards compatible to earlier generations. {\color{black}The wide ranging impact of the challenges discussed in the paper are summarized in Table~\ref{Tab:ChallengesSummary}.}

\begin{table*}[!t]
\centering
\scalebox{0.6}{
{\color{black}
\begin{tabular}{cccccc}
\toprule 
\textbf{Challenge (\# in the Paper)} & \hspace{10pt}\textbf{6G System Entities} & & & & \textbf{Impact on the Overall 6G System}
\tabularnewline
\midrule
\midrule 
&\textbf{Devices (UEs)} & \textbf{Base Station (BS)} & \textbf{Transport Network} & \textbf{Core Network}  & 
\tabularnewline
\midrule
\midrule
\textbf{Carrier Bandwidths (\#1)} & Yes & Yes & & & \hspace{-17pt}\tabitem Numbers of carriers to be aggregated for carrier aggregation  \tabularnewline
& & & & & \tabitem Number of bands which can be integrated from a device and BS \tabularnewline
& & & & & (CA combinations), inclusive of contiguous and non-contiguous bands \tabularnewline
& & & & & \tabitem SNR degradation with increasing bandwidth and impact on link budget/range \tabularnewline
& & & & & \hspace{-5pt}\tabitem Maintaining gain and phase uniformity of transceiver front-ends \tabularnewline 
& & & & & both at the device and BS link ends \tabularnewline 
& & & & & \tabitem Calibration of antenna element radiation patterns across the carrier bandwidths \tabularnewline 
& & & & & \hspace{-10pt}\tabitem Phase noise properties of oscillators at the BS and device ends \tabularnewline\midrule
\textbf{Semiconductor} & Yes & Yes & & & \tabitem 3D integration and in-packaging capability of RF front-ends \tabularnewline 
\textbf{Technologies (\#2)} & & & & & \tabitem Co-existence of digital and analog computations on the same MMIC\tabularnewline 
& & & & &  \tabitem Ability to maintain high output powers and constant SNRs  \tabularnewline \midrule
\textbf{Radio Transceiver Design (\#3)} & Yes & Yes & & & \tabitem Receiver noise figure and transmitter efficiency degraded \tabularnewline 
& & & & & \tabitem Power generation and heat dissipation at the BS \tabularnewline 
& & & & & \tabitem PA non-linearities causing OOBE \tabularnewline 
& & & & & \tabitem Control of inter-modulation products to meet sharing conditions \tabularnewline
& & & & & \tabitem Generation of coherent, low noise LO signals for beamforming \tabularnewline 
& & & & & \tabitem Integration of substrate, on-chip/off-chip antennas \tabularnewline 
& & & & & \tabitem Precision in packaging and manufacturing \tabularnewline\midrule
\textbf{Achieving Tb/s Capacity (\#4)} & Yes & Yes & Yes & Yes & \tabitem Link budgets determining the coverage and transmit power limits \tabularnewline 
& & & & & \tabitem Antenna array and transceiver configuration at the BS, \tabularnewline 
& & & & & i.e., distributing the BS array into multiple sub-arrays or operating \tabularnewline 
& & & & & with single centralized array: 1. Distributed PAs with high back off \tabularnewline 
& & & & & powers influencing OOBE for co-existence. 2. Backhaul \tabularnewline 
& & & & & infrastructure and latency/bandwidth constraints from distributed \tabularnewline 
& & & & & sub-arrays to the centralized unit. 3. Synchronization of backhaul infrastructure \tabularnewline 
& & & & & \tabitem Influence on initial access and beam management \\ 
& & & & & protocols operating at layers 1, 2 and 3 of the system\tabularnewline\midrule 
\textbf{Achieving Sub-ms Latencies (\#5)} & Yes & Yes & Yes & Yes & \tabitem Re-architecting the transport network to reduce transport latency \tabularnewline 
& & & & & \tabitem Introduce higher sub-carrier spacing to reduce RAN latency \tabularnewline 
& & & & & and its impact on cyclic prefix, and in turn cell range \tabularnewline 
& & & & & \tabitem TDD frame structures needed to reduce uplink acknowledgment delay \tabularnewline 
& & & & & and co-existence with other existing frame structures\tabularnewline 
& & & & & \tabitem Introduce MEC data centers at the wireless edge \tabularnewline 
& & & & & \tabitem Bringing user plane functions (UPFs) closer to the wireless edge \tabularnewline \midrule 
\textbf{Backwards Compatibility (\#6)} & Yes & Yes & & Yes & \tabitem CP-OFDM chosen as 5G waveform, and will \tabularnewline & & & & & limit the waveforms 6G can select \tabularnewline  
& & & & & \tabitem Spectrum co-existence compatibility for some 6G services \tabularnewline 
& & & & & \tabitem 6G operations first in limited geographies for\tabularnewline 
& & & & & seamless customer experience (as for 4G to 5G) \tabularnewline 
& & & & & \tabitem Dual connectivity support with 5G and 4G \tabularnewline 
& & & & & \tabitem 6GC must be backwards compatible with 5GC \tabularnewline\hline\bottomrule
\end{tabular}}}
\vspace{10pt}
\caption{{\color{black}Overall impact of the six critical challenges on the entire 6G system.}}
\vspace{-20pt}
\label{Tab:ChallengesSummary}
\end{table*}





\ifCLASSOPTIONcaptionsoff
  \newpage
\fi



%

\balance
%


\begin{IEEEbiographynophoto}{Harsh Tataria}
received a B.E. in Electronic and Computer Systems Engineering, and a Ph.D. in Communications Engineering from the Victoria University of Wellington, New Zealand, in December 2013 and March 2017. Since then, he has held several academic positions in the UK, USA, and Sweden. He is now with Ericsson AB, Sweden, as a Senior Researcher specializing in radio system standardization and realization. 
\end{IEEEbiographynophoto}

\begin{IEEEbiographynophoto}{Mansoor Shafi} is a Fellow at Spark NZ, Wellington, New Zealand. He has published extensively in cellular communications and has contributed prolifically to wireless standards in the ITU-R as a New Zealand delegate for over 30 years. He also contributes to 3GPP RAN 4 standardization. He is an Adjunct Professor at Victoria University of Wellington and the University of Canterbury, New Zealand. He is a Life Fellow of the IEEE and winner of multiple IEEE awards. 
\end{IEEEbiographynophoto}

\begin{IEEEbiographynophoto}{Mischa Dohler} is a Chief Architect at Ericsson Inc., Silicon Valley, USA. He was a Professor in Wireless Communications at King's College London, driving cross-disciplinary research and innovation in technology, sciences and arts. He is a Fellow of the IEEE, IET, the Royal Academy of Engineering and the Royal Society of Arts. He is a serial entrepreneur; composer \& pianist with 5 albums on Spotify/iTunes; He is fluent in 6 languages and receives ample coverage by the media.
\end{IEEEbiographynophoto}

\begin{IEEEbiographynophoto}{Shu Sun} obtained her B.S. degree from Shanghai Jiao Tong University in 2012, and received the Ph.D. degree from New York University (NYU) in 2018. Shu was a Systems Engineer with Intel Corporation from July 2018 to August 2021. She is now with Shanghai Jio Tong University, China. She has many highly-cited publications on millimeter-wave wireless communications including three award-winning papers. She is a recipient of many other awards from the IEEE and the Marconi foundation.
\end{IEEEbiographynophoto}

\end{document}